\documentclass[twocolumn]{aastex631}
\usepackage{subfigure}
\newcommand{\angstrom}{\mbox{\normalfont\AA}}

\usepackage{amssymb,amsfonts,amsmath,times,graphicx}
\newcommand{\changeurlcolor}[1]{\hypersetup{urlcolor=#1}} 
\usepackage{color}

\shorttitle{The Astrophysical Journal, Volume 915, Number 2} \shortauthors{Qutub et al.}
\received{2021 February 15} \revised{2021 May 27} \accepted{2021 May 27}

\begin{document}
\title{Depolarization of MgH Solar Lines by Collisions with Hydrogen Atoms}

\author[0000-0002-8688-4921]{Saleh Qutub}
\affil{Astronomy \& Space Science Dept, Faculty of Science, King Abdulaziz University, Jeddah, Saudi Arabia}
\author[0000-0001-7642-3518]{Yulia Kalugina}
\affil{Department of Optics and Spectroscopy, Tomsk State University, 36 Lenin av., Tomsk 634050, Russia}
\affil{Institute of Spectroscopy, Russian Academy of Sciences, Fizicheskaya St. 5, 108840 Troitsk, Moscow, Russia}
\author[0000-0003-3549-885X]{Moncef Derouich}

\affil{Astronomy \& Space Science Dept, Faculty of Science, King Abdulaziz University, Jeddah, Saudi Arabia}

\begin{abstract}
Interpretations of the very rich second solar spectrum of the MgH molecule face serious problems owing to the complete lack of any information about rates of collisions between the MgH and hydrogen atoms. This work seeks to begin the process of filling this lacuna by providing, for the first time, quantum excitation, depolarization, and polarization transfer collisional rates of the MgH ground state $X^2\Sigma$. To achieve the goals of this work, potential energy surfaces are calculated and then are included in the Schr\"odinger equation to obtain the probabilities of
collisions and, thus, all collisional rates. Our rates are obtained for temperatures ranging from $T \!\!=$2000 K to $T \!\!=$15,000 K. Sophisticated genetic programming methods are adopted in order to fit all depolarization rates with useful analytical functions of two variables: the total molecular angular momentum and temperatures. We study the solar implications of our results, and we find that the $X^2\Sigma$ state of MgH is partially depolarized by isotropic
collisions with neutral hydrogen in its ground state $^2S$. Our findings show the limits of applicability of the widely used approximation in which the lower-level polarization is neglected.\vspace{.3cm}
\emph{Unified Astronomy Thesaurus concepts}: \href{http://astrothesaurus.org/uat/1476}{Solar physics (1476)}; \href{http://astrothesaurus.org/uat/1477}{Solar atmosphere (1477)}; \href{http://astrothesaurus.org/uat/1503}{Solar magnetic
fields (1503)}
\end{abstract}


\section{Introduction}
Linear polarization, formed by scattering of anisotropic radiation and measured by observing the limb of the Sun, is called second solar spectrum (SSS). Numerical simulations of the SSS, stimulated by current and future spectropolarimetric projects, have opened new windows especially into the field of the quiet Sun’s magnetism (e.g., Bellot Rubio \& Orozco Su\'{a}rez 2019). The preparation of these projects and their scientific exploitation require collisional molecular data to be included in the coupled set of the radiative transfer equations and the statistical equilibrium equations (SEEs) for modeling the formation of the SSS.

The interest of molecular spectral lines observed in the SSS is twofold: first, they are in general optically thin lines, which facilitates modeling the formation of their polarization since the radiative transfer problem is less complicated when the line is optically thin. Second, each molecular multiplet contains numerous lines with different magnetic sensitivities (i.e., with
sufficiently different Land\'{e} g-factors) in a narrow spectral window that allows a multiline determination of the magnetic field, a technique known as the ``differential Hanle effect'' (e.g., Berdyugina \& Fluri 2004; Asensio Ramos \& Trujillo Bueno 2005; Bommier et al. 2006).

Nevertheless, different analyses (e.g., Berdyugina \& Fluri 2004; Asensio Ramos \& Trujillo Bueno 2005; Bommier et al. 2006) have obtained a value of $ \sim $ 7-15 G for the photospheric
turbulent magnetic field, which is clearly different from the value obtained by analyzing observations of the line polarization of Sr I $\lambda$4607 \AA ~($ \sim $ 40 G; e.g., Derouich et al. 2006). This difference seems to be due to the fact that collisions were usually neglected in the case of molecules as the molecular collisional rates are completely unknown. Therefore, a better
understanding of the SSS of molecules, and consequently a more accurate determination of the solar magnetic field, requires a precise determination of molecular collisional depolarization and transfer of polarization rates.

In particular, the scattering polarization of MgH is one of the most prominent features of the SSS (e.g., Mohan Rao \& Rangarajan 1999; Gandorfer 2000; Faurobert \& Arnaud 2003;
Asensio Ramos \& Trujillo Bueno 2005; Bommier et al. 2006; Mili\'{c} \& Faurobert 2012). MgH polarized lines must be analyzed in a comprehensive way to uncover important mysteries of the SSS and to address controversies surrounding Hanle effect diagnostics of turbulent magnetic fields at
subtelescopic scales (e.g., Bellot Rubio \& Orozco Su\'{a}rez 2019). Interpretation of the MgH polarized lines is difficult and incomplete because the Hanle effect and the effect of isotropic
collisions are mixed in the same observable (the polarization state; Mohan Rao \& Rangarajan 1999; Asensio Ramos \& Trujillo Bueno 2005; Bommier et al. 2006).

Our intention in this work is to provide new (de-)excitation, depolarization, and polarization transfer rates for the MgH molecule in its ground state $X^2\Sigma$ owing to collisions with the
hydrogen atom, H. These rates are very important in SSS studies. Computations of quantum collisional rates occur in two steps:
(1) determination of potential energy surfaces (PESs) for interaction of MgH and H, and
(2) study of the collisional dynamics by solving the Schr\"odinger equation with these PESs.
Reliable PESs for the interaction between H($^2S$) and MgH($X^2\Sigma^+$) were obtained by Ben Abdallah et al. (2009).
A thorough theoretical investigation of interaction potentials was carried out there, and the surfaces were represented in terms of Legendre polynomials. As a confirmation of the result of Ben
Abdallah et al. (2009), we have performed additional calculation of the PESs of the MgH-H system with higher resolution. As we show below, our PESs are in very good
agreement with those of Ben Abdallah et al. (2009). Nevertheless, our PESs are more accurate for radial separation larger than 9 $a_0$.

The treatment of the collision dynamics was made possible thanks to the MOLSCAT code (Hutson \& Green 1994). The infinite-order-sudden (IOS) approximation is adopted to compute (de-)excitation, depolarization, and polarization transfer cross sections for kinetic energies ranging from 50 to 37000  cm$^{-1}$ and for the first 70 rotational levels. This allows us to calculate
the corresponding rates for temperatures between 2000 and 15,000 K~\footnote{The data can be found at \href{https://doi.org/10.5281/zenodo.4694455}{10.5281/zenodo.4694455}}. Sophisticated genetic programming (GP) codes are used to infer analytical expressions depending on the temperature and total molecular angular momentum by fitting our collisional data (see Derouich et al. 2015). From the GP expressions, one can obtain depolarization collisional rates with accuracy better than 5\%. We study in some detail the solar
implications of our results.

\section{Theoretical Background}
We study the effects of isotropic collision of the MgH in the $^2\Sigma^+$ state with the hydrogen atom in its ground state $^2S$. We describe the MgH levels in Hund’s limiting case (b). Molecular
quantum numbers are denoted by $j$ and $N$, where  $j$ is the total angular momentum and $N$ is the rotational angular momentum related to $j$ by $\vec{j} \!=\! \vec{N} \!+\! \vec{S}_{MgH}$ where 
$S_{MgH} \!=\! 1/2$ is the spin of MgH. Therefore, $j \!=\! N \!\pm\! 1/2$. The spin of the hydrogen is $S_{H} \!=\! 1/2$; thus, the collision results in producing a singlet state $^1A'$ with total spin $S_{tot} \!=\! 0$ and a triplet state $^3A'$ with $S_{tot} \!=\! 1$.

The SSS of MgH molecule is quantified by using the density matrix formalism expressed on the basis of irreducible tensor operators (ITOs), which has been introduced by Fano (1957) and then adopted in solar physics by many authors (e.g., Sahal-Br\'{e}chot 1977; Trujillo Bueno 2001; Landi Degl'Innocenti \& Landolfi 2004). In the ITOs basis, the density matrix elements are denoted by $ \rho_{q}^{k} (j) $ with a tensorial order $0  \! \leqslant \!  k  \! \leqslant \! 2 j$ and a coherence number $-k  \! \leqslant \!  q  \! \leqslant \! k$. The state of the radiation emitted by the MgH molecule can be obtained by knowing the $ \rho_{q}^{k} (j) $. In fact, intensity, circular polarization, and linear
polarization are associated with the $ \rho_{q}^{k} (j) $ elements of order
$k$ = 0, $k$ odd integer (i.e. $k$=1, 3, 5, etc.), and $k$ even integer  (i.e. $k$=2, 4, 6, etc.),   respectively.
The contribution of collisions to the evolution of the density
matrix $\rho$ is given by the following rate equations:
\begin{eqnarray} \label{eq_ch3_17}
\Big(\frac{d \; ^{j}\rho_q^{k}}{dt} \Big)_{coll} & \!=\! & - D^k(j, T) \; ^{j}\rho_q^k \nonumber \\
&& \!- ^{j}\rho_q^k \! \sum_{j' \ne j} \! \sqrt{\frac{2j'+1}{2j+1}} D^0 (j \!\to\!  j', T) \\
&& \!+\! \sum_{j' \ne j} \! 
D^k(j' \!\to\!  j, T) \;  ^{j'}\rho_q^k \,.  \nonumber 
\end{eqnarray}
$ D^k(j, T)$ are the  depolarization rates  of the  $j$-level   due to purely elastic collisions, 
 and  $ D^k(j \!\to\! j', T)$ are the rates of polarization transfer 
 between the  $j$ and $j'$ levels.
 
Note that apart from the multiplicity factor $\sqrt{(2j'\!+\!1)/(2j\!+\!1)}$, the $C^{k}(j,j')$ denoted by Landi Degl'Innocenti \& Landolfi (2004) become the collisional transfer rates $D^{k}(j'\!\to\!j)$ defined here and in Sahal-Br\'{e}chot (1977) and adopted by Derouich et al. papers (e.g.  Derouich et al.  2003  and Derouich  2006). One can  refer to  Derouich \& Ben Abdallah 2009 for more details about the origin of the multiplicity factor $\sqrt{(2j'\!+\!1)/(2j\!+\!1)}$. We emphasize that, after plugging in all collisional rates, the final collisional rate equations, $({d \; ^{j}\rho_q^{k}}/{dt} )_{coll}$, become exactly the same in both conventions.  The collisional rates are obtained through integration of cross-sections $\sigma^k$ over Maxwellian distribution of relative velocities  (e.g. Derouich  2006). In addition,
\begin{eqnarray} \label{gammak}
D^k(j,T) \!=\! D^0(j \!\to\! j,T) \!-\! D^k(j \!\to\! j,T),
\end{eqnarray}
which implies that $D^0(j) \!=\! 0$. 

 We use the approach of  Corey \& Alexander (1985)  and  Corey et al. (1986)  to obtain  expressions  for  the  polarization transfer cross-sections $\sigma^k(j \!\to\! j',E)$  and depolarization cross-sections $\sigma^k(j,E)$. In addition,   the  IOS approximation is adopted which can be well justified especially for sufficiently high temperatures (see e.g. Lique et al. 2007).   In these conditions, the $\sigma^k$ adopted in this work are given, for example, by Eq.~(1) of Qutub et al. (2020).
The total collisional rates averaged over spin can then be calculated via the relation (Corey \& Alexander 1985):
\begin{eqnarray} \label{eq:TotalDepolarization}
D^k(j \to j',T) &=&  \frac{1}{4} \, \big[3 \, D^k(j \to j',T; \; ^3A')
\nonumber \\
&& +  D^k(j \to j',T; \;  ^1A') \big] \,.
\end{eqnarray}
%
%

\section{Potential Energy Surfaces} \label{sec:PES}
\begin{figure*}[ht]
\centering
\includegraphics[width=8.9cm]{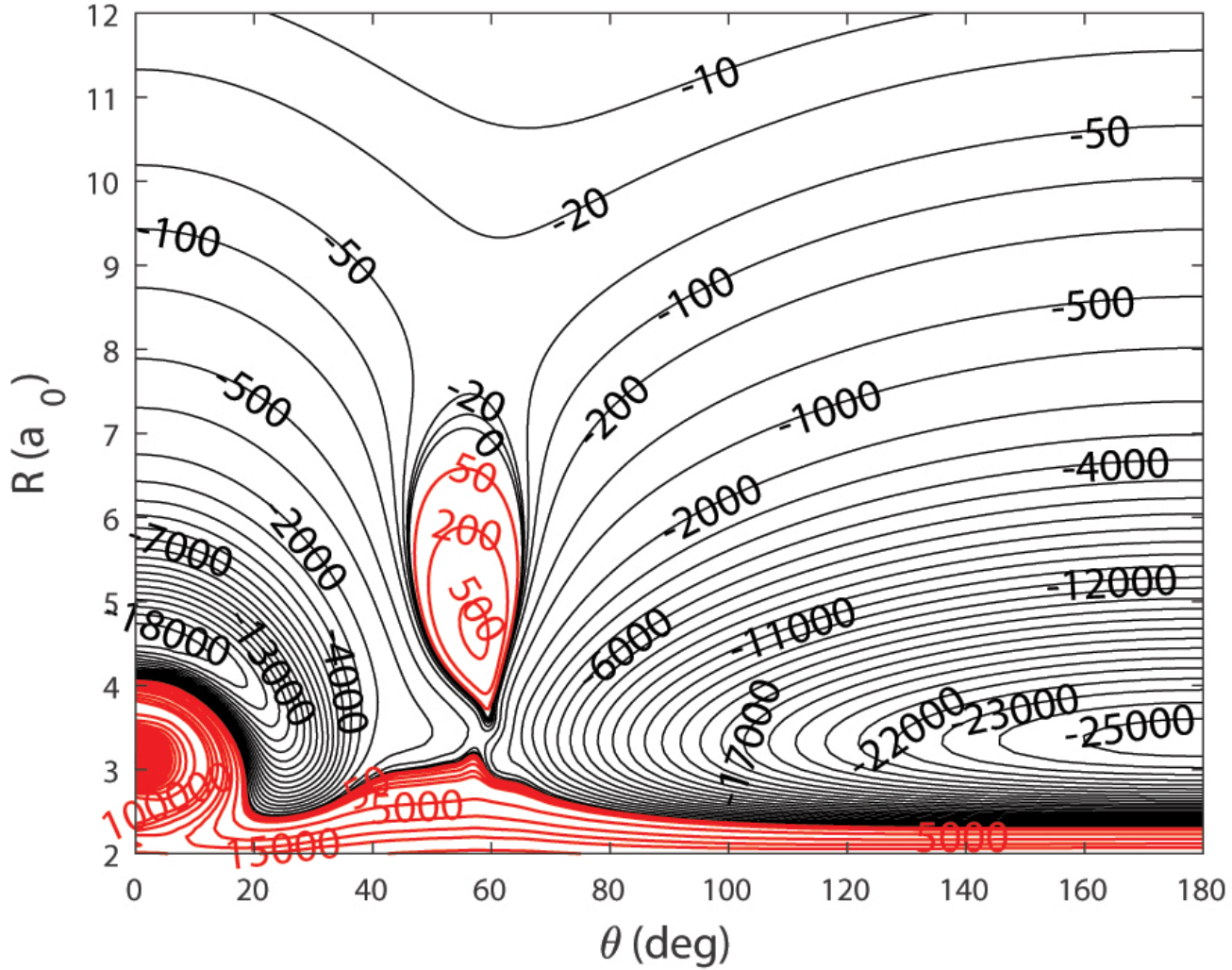}
\includegraphics[width=8.9cm]{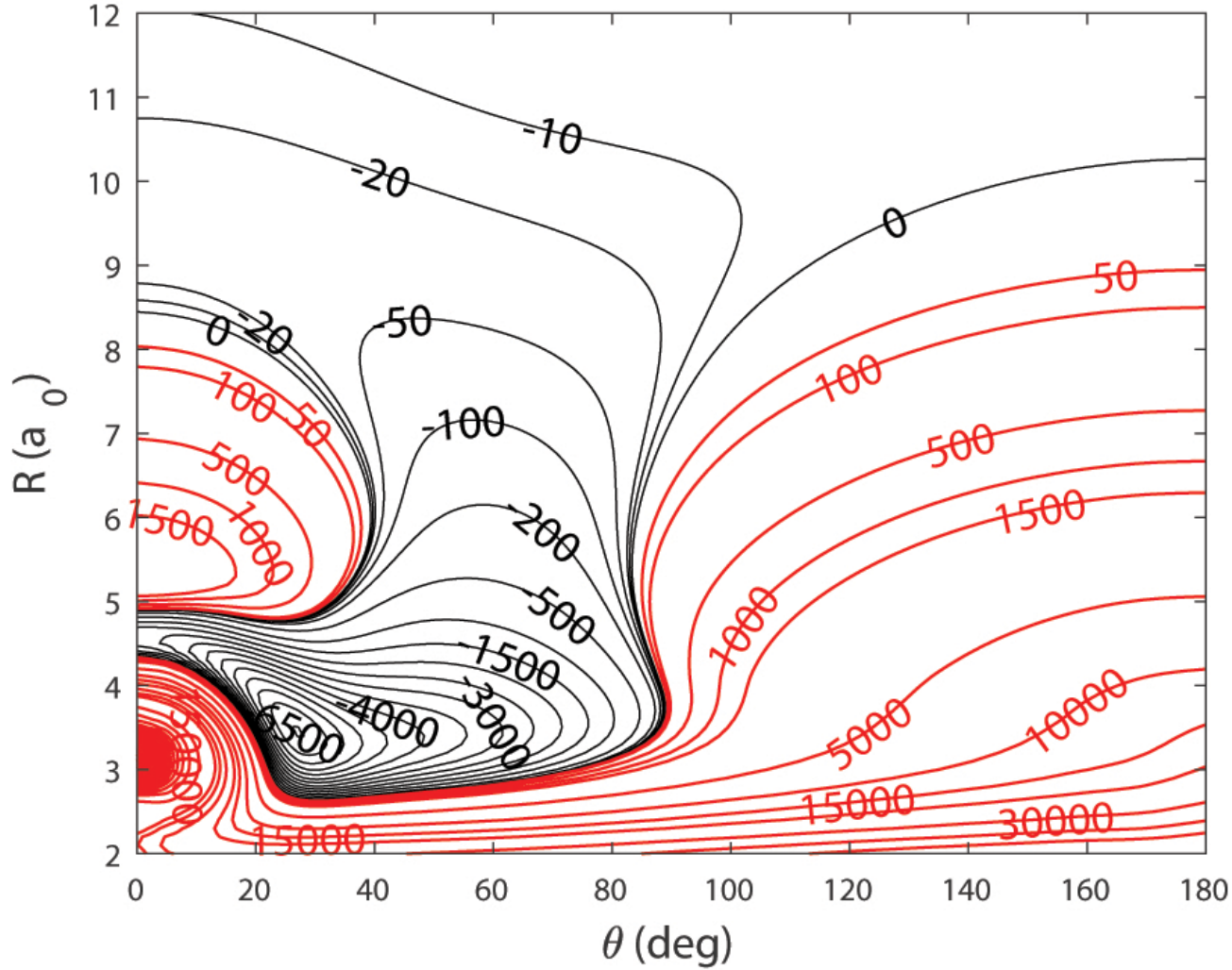}
\caption{Two-dimensional PES for  $^1A'$ state (left panel) and $^3A'$ state (right panel). Energy is in cm$^{-1}$.}   
\label{fig:pes}
\end{figure*}
We adopt the coordinate system of Jacobi ($R$, $r_{MgH}$, $\theta$) for the calculation of PESs. Here $R$ represents the distance from the center of mass of the MgH molecule to the H atom,
and $\theta$ is the rotation angle of the H atom around the MgH. The MgH molecule is assumed to be rigid with Mg-H distance frozen at its equilibrium value $r_{MgH}$~=~3.2692 ~$a_0$ (Rosen 1970).

Accurate ab initio computations of the PESs for the $^1A'$ and $^3A'$ states are performed in the internally contracted multireference configuration interaction level of theory (Werner \& Knowles 1988). Partial size consistency is corrected by following the Davidson (+Q) correction (Davidson \& Silver 1977). The remaining correction is made by subtracting the energy at $R$=100 $a_0$. The
five lowest orbitals of the Mg atom were kept frozen. The active space consists of four electrons distributed in six active orbitals. The augmented correlation-consistent triple zeta (V5Z) basis set (Dunning 1989) for Mg and the VQZ basis set for the H atoms were used. All the PESs are obtained using the MOLPRO package (e.g., Werner et al. 2010).

For the $^1A'$ state the $R$ values were varied from 1.75 to 50 $a_0$, giving 55 grid points. For the $^3A'$ state the $R$ values were varied from 2.0 to 50 $a_0$, with a total of 61 grid points.
We used a variable step in angle $\theta$ in order to cover the complex behavior of both PESs. The total number of ab initio points is 3300 for the singlet state and 3294 for the triplet state. We checked the energy convergence for more problematic regions (0$^\circ$--20$^\circ$ and 150$^\circ$--180$^\circ$) by taking different starting points for ab initio calculations. For $^1A'$ and $^3A'$ potentials, the 2D spline was employed. This allows us to avoid fitting errors.

The resulting PESs for the $^1A'$ and $^3A'$ electronic states are shown in Figure~\ref{fig:pes}. For the singlet state, there are two minima on the PES associated with the formation of HMgH and MgHH molecules. The HMgH arrangement corresponds to the minimal structure with $\theta\!=\!180^{\circ}$ and $R \!=\!3.36$ $a_0$ and has the well depth $E \!=\! - 25531.5$ cm$^{-1}$. The minimum compares well with the $E \!=\! - 25561.55$ cm$^{-1}$ at $R \!=\!3.40$ $a_0$ obtained by Ben Abdallah et al. (2009). The MgHH minimal structure corresponds to $\theta \!=\! 0^{\circ}$ and $R=$4.59 $a_0$ and has a well depth $E\!=\!- 18791.2$ cm$^{-1}$ (compared to $E\!=\!- 19642.06$ cm$^{-1}$ at $R \!=\!4.60$ $a_0$ of Ben Abdallah et al. 2009). The minimum for the triplet state occurs at $R\!=\!3.45$ $a_0$, $\theta\!=\!26^{\circ}\!\!.99$ and has an energy $E\!=\!- 6531.3$ cm$^{-1}$
(compared to $E\!=\! - 6758.80$ cm$^{-1}$ at $R$=3.2 $a_0$ and $\theta\!=\!33^{\circ}$ of Ben Abdallah et al. 2009).

\section{Results and Discussions}
%
%
\subsection{Depolarization Rates} \label{sec:Depolarization}
\begin{figure*}[ht]
\centering
\includegraphics[width=8.9cm]{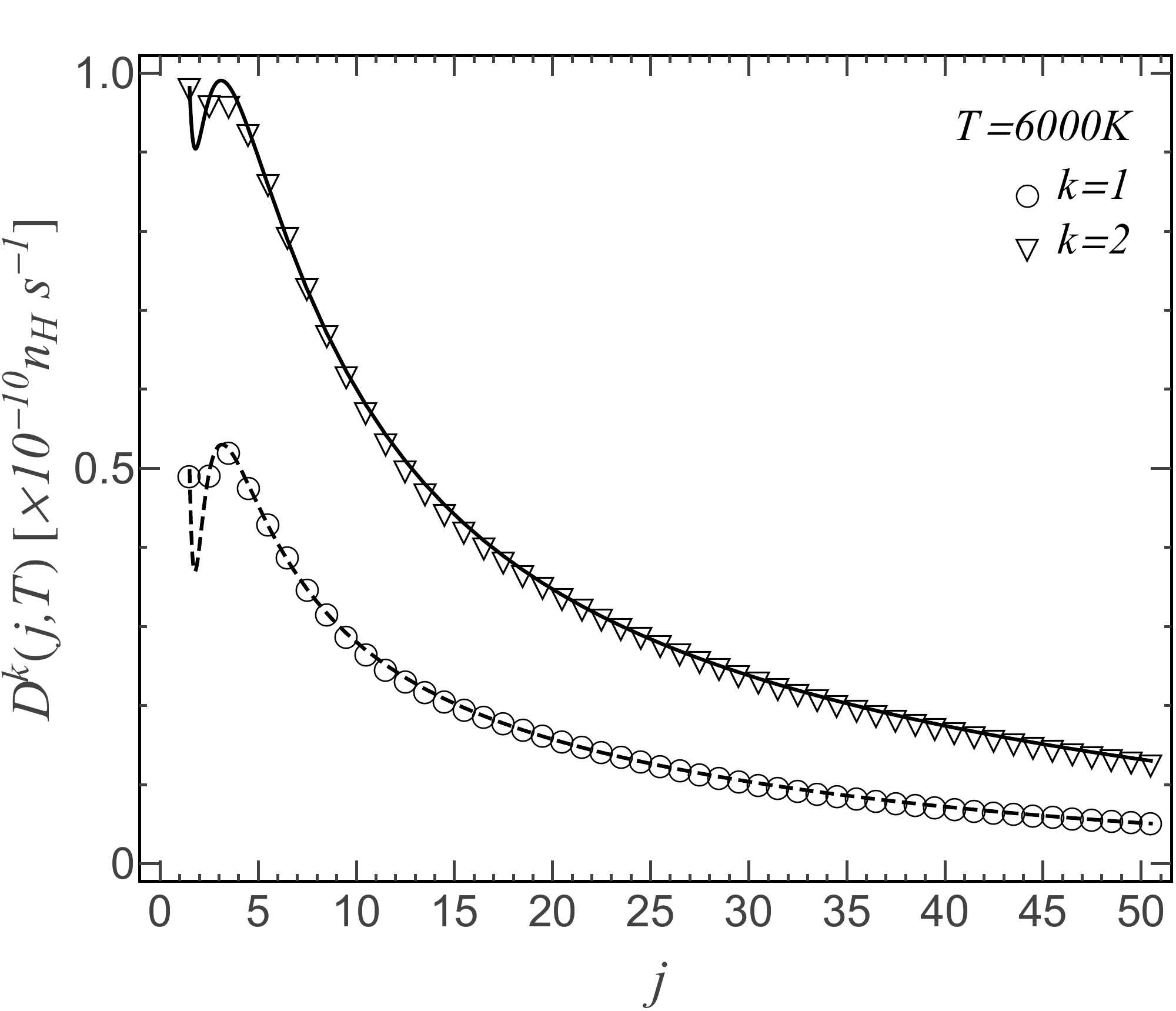} 
\includegraphics[width=8.9cm]{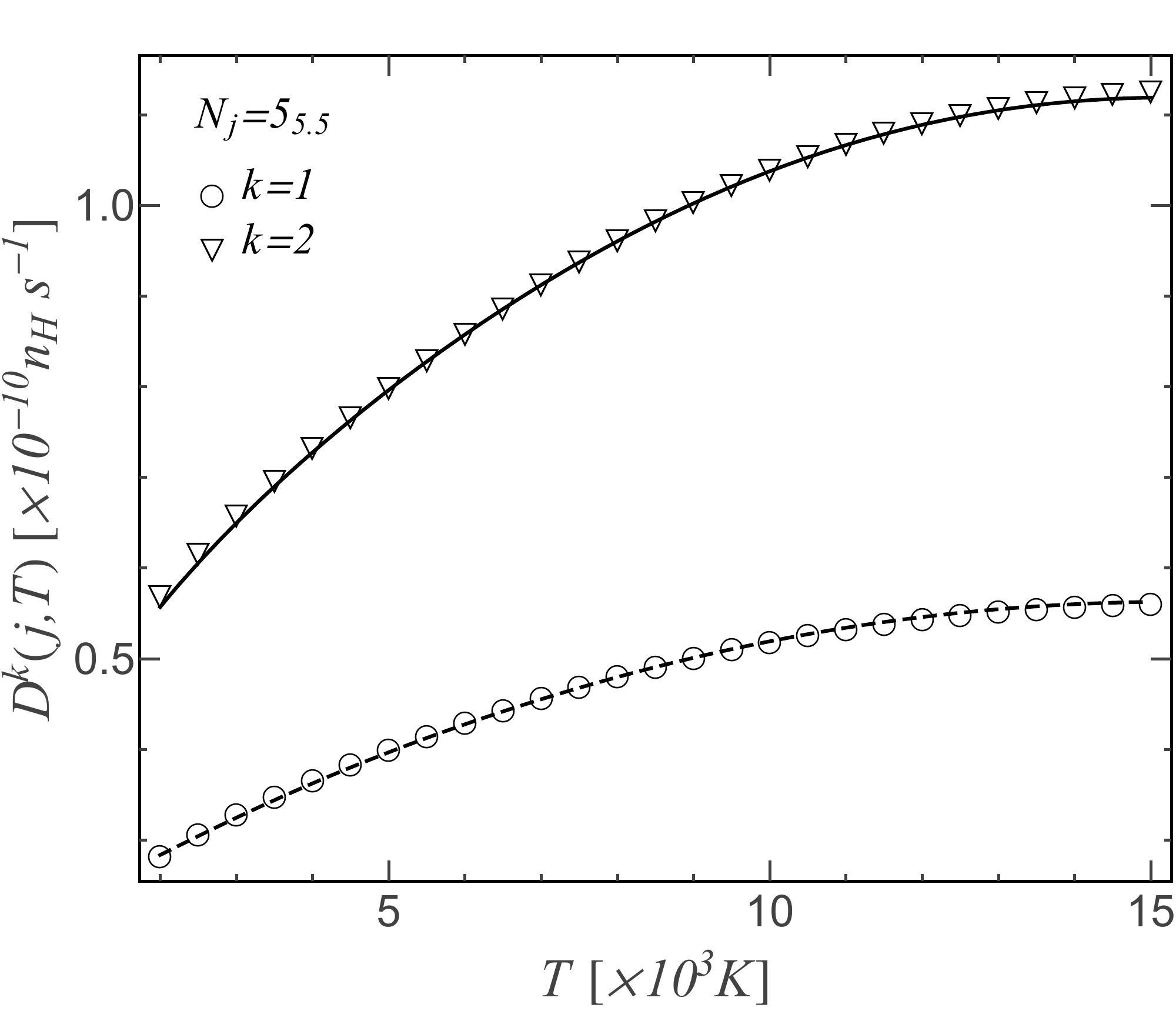}
\caption{Variation with $j$ (left panel) and with $T$ (right panel) of the collisional depolarization rates for $k \!=\! 1$ (open circles) and $k \!=\! 2$ (open triangles). The dashed and solid curves show the GP fit values obtained using Equations~(\ref{eq:depolk1}) and (\ref{eq:depolk2}), respectively.} 
\label{fig:Pol_jT}
\end{figure*}  
In Figure~\ref{fig:Pol_jT}, we show the variation of collisional depolarization rates for the orientation, $k\!=\! 1$ (open circles), and the alignment, $k\!=\! 2$ (open triangles), as a function of $j$ at $T\!=\! 6000$ K in the left panel and as a function of $T$ for the level
$N_j\!=\!5_{5.5}$ in the right panel. As one would expect, the collisional depolarization rates increase with temperature (roughly $D^1,D^2 \!\propto\! T^{0.34}$ for the given level) and decrease with increasing $j$ (roughly $D^1 \!\propto\! j^{-0.78}$ and $D^2 \!\propto\! j^{-0.70}$ for the given temperature) as the energy separation between rotational levels decreases with increasing $j$.

It is clear from Figure~\ref{fig:Pol_jT} that the depolarization rates with tensorial order $k\!=\!2$ are larger than those with tensorial order $k\!=\!1$. Using GP fitting techniques, we obtain the following relations, which represent the depolarization rates in the temperature range 2000 -- 15,000 K and for total angular
momentum up to 50.5 with error less than 5\%:\footnote{Separate fits for the singlet and triplet contributions are also available from the authors upon request.}
\begin{widetext}
\begin{eqnarray} \label{eq:depolk1}
&&\frac{D^{1}(j,T)}{n_{\rm H} \!\!\times\!\! 10^{-10}} \!\!=\!\! 
\frac{  0.0004582 j^{3.9722} T^{0.41185} \!\!-\! 0.0005562 j^{3.9806} T^{0.398} \!\!+\! 1.23653 j^{0.19053} T^{0.000014} \!\!-\! 10.4679 j^{0.02168} \!\!+\!  9.2281}
   { 15.5575 \frac{j^{4.52315}}{T^{0.518178}} \!\!-\! 1.58 \!\times\! 10^{-8} j^{6.79} T^{0.326} \!\!-\! 7.2 \!\times\! 10^{-16} j^{4.786} T^{2.69} \!\!-\! 15.2825 \frac{j^{4.5256}}{T^{0.51576}} \!\!-\! \frac{0.024}{T^{0.36}}} ,
\\
\label{eq:depolk2}
&& \frac{D^{2}(j,T)}{n_{\rm H} \!\!\times\!\! 10^{-10}} \!\!=\!\! 
\frac{0.0005 \!+\! 15.5158 \frac{T^{0.00011}}{j^{1.5021}} \!\!+\! \frac{0.048}{j^{3.26}
   T^{0.014}} \!\!-\! \frac{15.56}{j^{1.50455}}}
   { 7  \!\times\! 10^{-10 } j^{2.634} T^{0.49} \!\!+\! 0.046 \frac{j^{1.045}}{T^{0.84}} \!\!+\! 4.516 \!\times\! 10^{-7}  \frac{T^{1.1984}}{j^{0.0017}} \!\!+\! \frac{0.0352}{j^{0.079} T^{0.22}} \!\!-\! 6.544 \!\times\! 10^{-7} T^{1.161}} .
\end{eqnarray}
\end{widetext}
The dashed and solid curves in Figure~\ref{fig:Pol_jT} represent the GP fit values calculated using Equations~(\ref{eq:depolk1}) and (\ref{eq:depolk2}), respectively, which are in very good agreement with the directly calculated rates.

\subsection{(De-)excitation and Transfer of Polarization Rates} \label{sec:PolarizationTransfer}
\begin{figure*}[ht]
\centering
\includegraphics[width=8.9cm]{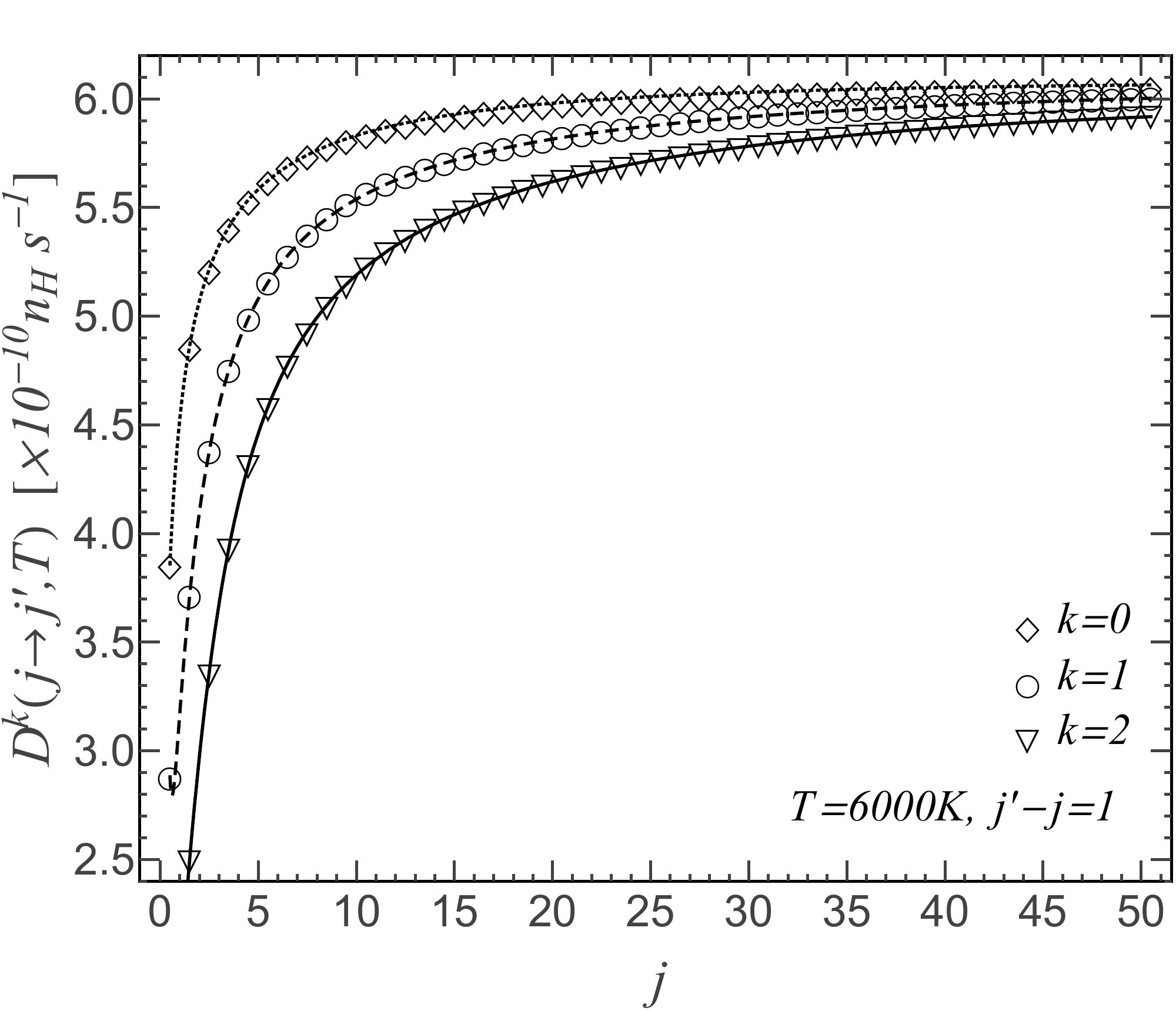}
\includegraphics[width=8.9cm]{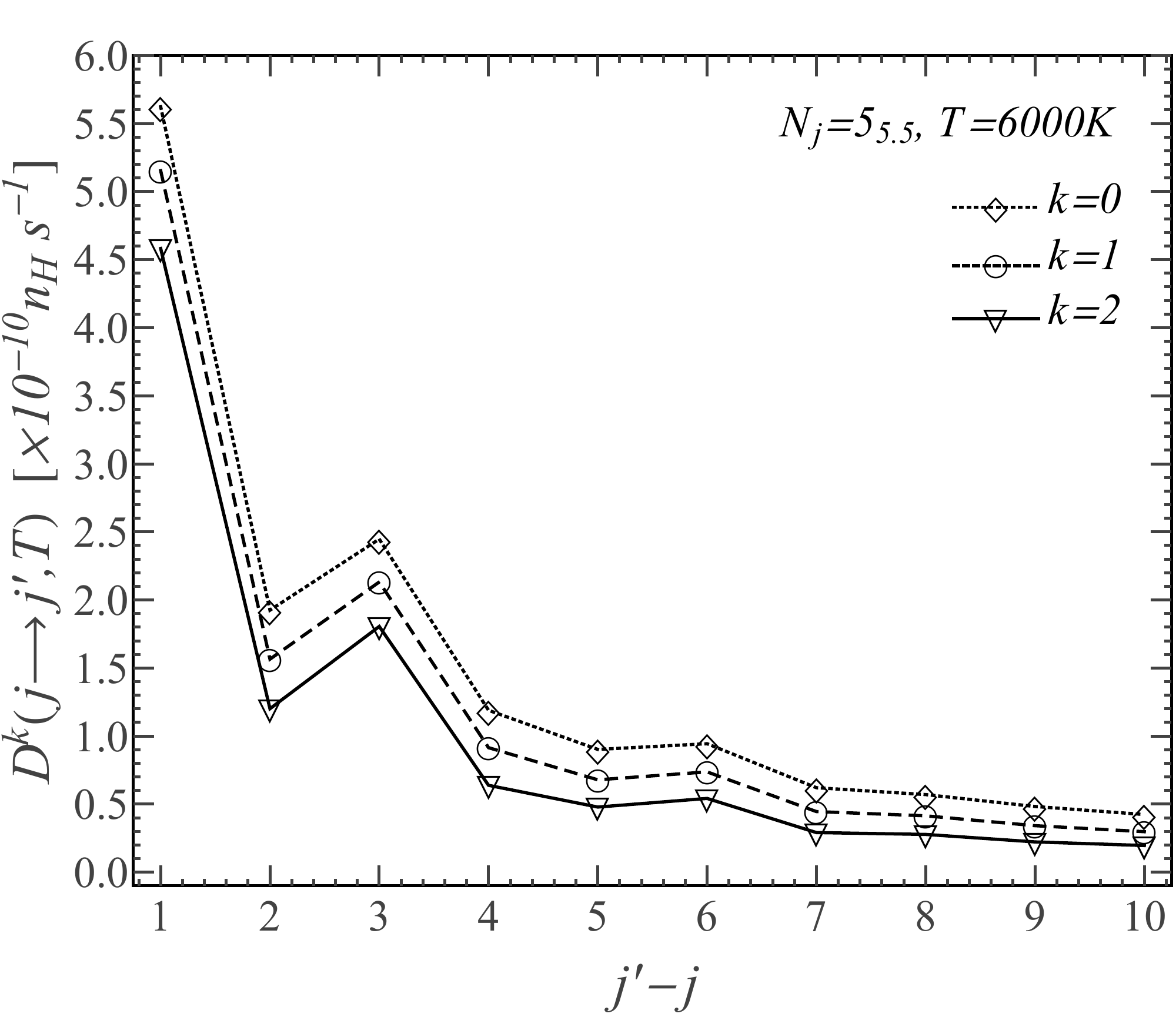} 
\caption{Variation of the collisional transfer rates   for $k\!=\!0$, $k\!=\!1$, and $k=2$  as functions of $j$ (left panel) for $j'\!-\!j\!=\!1$ and $T\!=\!6000$ K and as functions of $j'\!-\!j$ for the level $N_j\!=\!5_{5.5}$ and $T\!=\!6000$ K. The dotted, dashed, and solid curves in the left panel show the GP fit values obtained using Equations~(\ref{eq:Transk0})--(\ref{eq:Transk2}), respectively.} 
\label{fig:Trans_jdeltj}
\end{figure*}  
We now turn our attention to the (de-)excitation and polarization transfer rates. In Figure~\ref{fig:Trans_jdeltj}, we show the variation with $j$ in the left panel (for $j'\!-\!j\!=\!1$ and $T\!=\!6000$ K) and with $j'\!-\!j$ in the right panel (for $N_j\!=\!5.5$ and $T\!=\!6000$ K) of the upward transfer of population $k\!=\!0$ and of polarization $k\!=\!1,2$ collisional rates. Note that the collisional (de-)excitations rates, $C(j \!\to\! j')$, are related to the rates of transfer of population due to collisions, $D^0(j \!\to\! j')$, via the relation $C(j \!\to\! j') \!=\! \sqrt{(2j'\!+\!1)/(2j\!+\!1)}\, D^0(j \!\to\! j')$ (e.g. Derouich 2006).
One can see from the left panel of Figure~\ref{fig:Trans_jdeltj} that the transfer rates increase with increasing $j$ as the energy difference between levels decreases with $j$: $D^0(j\!\to\! j\!+\!1) \!\propto\! j^{-0.11}$, $D^1(j \!\to\! j\!+\!1) \!\propto\! j^{-0.18}$, and $D^2(j \!\to\! j\!+\!1) \!\propto\! j^{-0.37}$ roughly upto $j\!=\!15$ for the case at hand. For the same reason the collisional transfer rates decrease with increasing $\vert j'\!-\!j \vert$ (see the right panel of Figure~\ref{fig:Trans_jdeltj}): roughly $D^0(j \!\to\!j') \!\propto\! \vert j'\!-\!j \vert^{-1.1}$, $D^1(j \!\to\!j') \!\propto\! \vert j'\!-\!j \vert^{-1.2}$, and $D^2(j \!\to\!j') \!\propto\! \vert j'\!-\!j \vert^{-1.3}$ for the given case. As one would expect, the collisional transfer rate with $\vert j'\!-\!j \vert \!=\!1$ are dominant, as can be seen from the right panel of Figure~\ref{fig:Trans_jdeltj}. Therefore, by using GP fitting techniques, we obtain the following relations, which represent the collisional transfer rates with $j'\!-\!j\!=\!1$ in the temperature range 2000 -- 15,000 K and for total angular momenta up to 50.5 with maximum error less than 1\%:\footnote{Separate fits for the singlet and triplet collisional transfer rates, in addition to fits for the collisional transfer rates with $ j'\!-\!j \!>\!1$, are available from the authors upon request.}
\begin{widetext}
\begin{eqnarray} \label{eq:Transk0}
&& \hspace{-1.cm} \frac{D^0(j \!\to\! j\!+\! 1,T)}{n_{\rm H} \!\!\times\!\! 10^{-10}} \!=\!
\frac{1.82599342 j^{0.72733479} T^{0.000035887} \!\!+\! 2.23672898 \frac{j^{0.72827942}}{T^{0.000028873}} \!\!+\! 2129.2 \frac{j^{0.3877}}{T^{1.9607}} \!\!-\! 4.06270986 j^{0.727856143} \!\!-\! 6.4 \!\times\! 10^{-6}}
   {3.42 \!\times\! 10^{-9} j^{0.389} T^{0.662} \!\!+\! 0.02708 \frac{j^{0.19}}{T^{0.9462}} \!\!+\! \frac{825}{j^{0.252} T^{2.1375}} \!\!+\! 6680 \frac{j^{0.402}}{T^{2.311}} \!\!-\! \frac{0.0563}{T^{1.051}}} ,
\nonumber \\ 
\hspace{-2.cm} && \hspace{-2.cm}
\\ 
\label{eq:Transk1}
&& \hspace{-1.cm}  \frac{D^1(j \!\to\! j\!+\! 1,T)}{n_{\rm H} \!\!\times\!\! 10^{-10}} \!=\!
\frac{317.35 \frac{j^{2.2305}}{T^{0.855}} \!\!+\! 2.118 \frac{j^{1.2645}}{T^{0.03446}} \!\!+\! 3.347 \!\times\! 10^{-9} j^{2.1953} T^{1.812}  \!\!-\! 2.619 j^{0.2628} \!\!+\! 1.8358  }
   { 1.56 \!\times\! 10^{-11} j^{1.414} T^{2.2627} \!\!+\! 1.35 \!\times\! 10^{-10} j^{2.2524} T^{1.9059} \!\!+\! 21 \frac{ j^{1.2844}}{T^{0.4977}} \!\!+\! 959.7 \frac{j^{2.2353}}{T^{1.19094}} \!\!-\! \frac{332109}{T^{2.31}}} ,
\\
\label{eq:Transk2}
&& \hspace{-1.cm}  \frac{D^2(j \!\to\! j\!+\! 1,T)}{n_{\rm H} \!\!\times\!\! 10^{-10}} \!=\!
\frac{ 3.16488 j^{3.08986} T^{0.000001128} \!\!+\! 3.18179 j^{0.000029} T^{0.00000074} \!\!-\! 3.16486 j^{3.08986}  \!\!-\! 0.002713 \frac{j^{2.8955}}{T^{0.4695}}  \!\!-\! 3.18176}
     { 1.52 \!\times\! 10^{-15} j^{2.849} T^{2.206} \!\!+\! 0.002613 \frac{j^{2.8175}}{T^{0.5484}} \!\!+\! 0.00521 \frac{j^{1.076}}{T^{0.503}} \!\!+\! 4.6 \!\times\! 10^{-17} j^{1.066} T^{2.677}  \!\!-\! \frac{0.21}{T^{1.165}}} .
\end{eqnarray}
\end{widetext}
The dotted, dashed, and solid curves in the left panel of Figure~\ref{fig:Trans_jdeltj} represent the GP transfer rates calculated using Equations~(\ref{eq:Transk0})--(\ref{eq:Transk2}), respectively, which agree extremely well with the original rates. The GP analytical functions given in Equations~(\ref{eq:Transk0})--(\ref{eq:Transk2}) can be implemented in the numerical codes calculating the theoretical polarization to generate the rates for any $j$ and $T$ values.

We remark that the collisional transfer rates have similar behavior with temperature to the collisional depolarization rates. Downward collisional transfer rates exhibit a similar behavior with $j$ and $T$ to the upward transfer rates. In fact, for isotropic collisions, which is the case under consideration, one has (e.g., Derouich et al. 2007)
\begin{eqnarray} 
D^k(j_u \!\to\! j_\ell,T) \!=\!
\frac{2 j_\ell \!+\! 1}{2 j_u \!+\! 1} \exp \!\left(\! \frac{\Delta E_{j_u,j_\ell}}{k_{\rm B} T}  \!\right)\! D^{k}(j_\ell \!\to\! j_u, T)  ,  \nonumber  
\end{eqnarray}
with $\Delta E_{j_u,j_\ell}$ being the energy difference between the upper $j_u$ and lower $j_\ell$ levels and $k_{\rm B}$ being the Boltzmann constant.

It is interesting to note that for the MgH molecule and other molecules such as the CN (Qutub et al. 2020), collisional depolarization rates are significantly smaller than collisional transfer rates. It is also worth noting that the collisional depolarization and transfer rates are larger for the MgH molecule as compared to those of the CN molecule. This is due to the MgH molecule being more asymmetric than the CN molecule.

\subsection{On the Accuracy of the Collisional Rates}
There are no experimental or other theoretical values of depolarization and polarization transfer rates associated with MgH solar lines to compare with. In addition, as far as we know, neither experimental nor theoretical work is being currently performed to provide detailed collisional data that would enable a quantitative analysis of the MgH polarization. This work is a first step toward complete determination of the MgH depolarization and polarization transfer rates.

We use up-to-date quantum methods to calculate new PESs and to solve the collision dynamics allowing the calculation of the rate coefficients. The IOS approximation (e.g., Goldflam et al. 1977; Parker \& Pack 1978) used in this work is known to be sufficiently precise for solar temperatures (e.g., Derouich \&
Ben Abdallah 2009). Our quantum collisional rates should be sufficiently accurate for solar applications.

\subsection{Solar Implications}  \label{sec:SolarImplications}
\begin{table*}[ht]
\centering 
\caption{Comparison of the Inverse Lifetime $\frac{1}{t_{life}}$=$B_{\ell u} I(\lambda)$ of the MgH~$X^{2}\Sigma$ State to Its Linear Depolarization Rates $D^2$}  
\begin{tabular}{c c c c c c c c c c} 
\hline
\hline
$\lambda_{u \ell} \, (\angstrom)$ & $N_{\ell}$ & $j_{\ell}$ & $I(\lambda_{u \ell}) \, (10^{-5} {\rm erg} $ & $A_{u \ell}$ & $B_{\ell u} I(\lambda_{u \ell})$ &
\multicolumn{2}{c}{$\omega_{L} \vert g_{j_{\ell}} \vert \, (10^{7} {\rm s}^{-1})$}  & 
\multicolumn{2}{c}{$D^{2}(N_{\ell} j_{\ell},T \!=\! 5778~\rm{K}) \ (10^{5} {\rm s}^{-1})$}
\\
  &   &   &   ${\rm cm}^{-2} \,{\rm s}^{-1} {\rm sr}^{-1} {\rm Hz}^{-1})$  &  $(10^{7}{\rm s}^{-1})$  & $(10^{5} {\rm s}^{-1})$ & ${\rm B} \!=\! 10 {\rm G}$ & ${\rm B} \!=\! 100 {\rm G}$ &  $n_{\rm H} \!=\! 10^{15} {\rm cm}^{-3}$  &  $n_{\rm H} \!=\! 10^{16} {\rm cm}^{-3}$
\\
\hline
$5170.574$ & 12 & 11.5 & $2.84645$ & $1.99441$ & $1.97524$ & $0.70353$ & $7.03530$ & $0.52392$ & $5.23920$
\\
$5171.012$ & 12 & 12.5 & $2.75250$ & $1.99977$ & $1.91567$ & $0.70353$ & $7.03530$ & $0.49026$ & $4.90260$
\\
$5174.895$ & 10 & 9.5 &	 $2.89082$ & $1.98276$ & $1.99932$ & $0.83753$ & $8.37530$ & $0.60793$ & $6.07930$
\\
$5175.419$ & 10 & 10.5 & $2.91406$ &	$1.98974$ & $2.02310$ &	$0.83753$ & $8.37530$ & $0.56288$ & $5.62880$
\\
$5176.816$ & 9 & 8.5 & $2.88194$ & $1.97566$ & $1.98825$ & $0.92569$ & $9.25690$ & $0.65950$ & $6.59500$
\\
$5178.503$ & 8 & 7.5 & $2.94770$ & $1.97052$ & $2.03031$ & $1.03460$ & $10.34600$ & $0.71797$ & $7.17970$
\\
$5179.994$ & 7 & 6.5 & $1.54633$ & $1.95580$ & $1.05803$ &	$1.17255$ & $11.72550$ & $0.78215$ & $7.82150$
\\
$5180.593$ & 7 & 7.5 & $2.87995$ & $1.96892$ & $1.98444$ & $1.17255$ & $11.72550$ & $0.71797$ & $7.17970$
\\
$5181.307$ & 6 & 5.5 & $2.31560$ & $1.93665$ & $1.57007$ & $1.35294$ & $13.52940$ & $0.84786$ & $8.47860$
\\
$5181.930$ & 6 & 6.5 & $2.60932$ & $1.95880$ & $1.79011$ &	$1.35294$ & $13.52940$ & $0.78215$ & $7.82150$
\\
\hline
\hline
\end{tabular}
\begin{flushleft}
\textbf{Note}. Also compared is $B_{\ell u} I(\lambda)$ with the values $(\omega_L \vert g_{j_{\ell}} \vert)^{-1}$ that estimate the Hanle depolarization.
\end{flushleft}
\label{tab:sol_impli}
\end{table*}
Let us briefly highlight the importance of our collisional rates for solar spectropolarimetry. Rotational levels of the electronic ground state of the solar MgH molecule, $X^{2}\Sigma$, can be polarized owing to the anisotropy of the incident radiation. This polarization could either be transferred to the MgH upper electronic states via radiative absorption, hence contributing to polarization of the emitted radiation, or get destroyed by isotropic collisions. This is usually quantified by solving the full set of coupled SEEs governing the population and polarization of different atomic or molecular levels taking into account all the intervening processes. However, this goes beyond the scope of this work.

Nevertheless, for the purpose of exploring the possible effect of collisions on the MgH ground-state depolarization, it is sufficient to compare the radiative transfer rates due to absorption for the rotational levels of the MgH electronic ground state, $B_{\ell u} I(\lambda_{u \ell})$ (which determine lifetimes of the levels of the electronic ground state, $t_{\rm life}^{-1} \!=\! B_{\ell u} I(\lambda_{u \ell})$), with the corresponding collisional depolarization rates, $D^k(j_\ell)$. Here $I(\lambda_{u \ell})$ denotes the intensity of light of wavelength $\lambda_{u \ell}$ at the center of the solar disk incident on the MgH molecules, and $B_{\ell u} \!=\! (g_u/g_{\ell}) (c^2/2h\nu_{u \ell}^3) A_{u \ell}$ is the Einstein absorption coefficient, with $A_{u \ell}$ being the Einstein coefficient for spontaneous emission, $g_u$ and $g_{\ell}$ the multiplicity of upper and lower levels, $h$ Planck’s constant, and $c$ the speed of light. For concreteness, we contrast the collisional linear depolarization rates of the state $X^{2}\Sigma$, $D^{2}(j_{\ell})$, calculated at the effective photospheric temperature, $T\!=\! 5778$~K, and for the typical photospheric density of hydrogen: $n_{H}  \!=\!  10^{15} \!-\! 10^{16}$ cm$^{-3}$, with the corresponding radiative absorption rates, $B_{\ell u} I(\lambda_{u \ell})$, for some representative lines of the $A^{2}\Pi-X^{2}\Sigma$ system of MgH.

We display the values of both $D^{2}(j_{\ell})$ and $B_{\ell u} I(\lambda_{u \ell})$ for
selected lines in Table~\ref{tab:sol_impli}. The values of the core relative intensity of the selected lines were obtained from Delbouille et al. (1972), and the corresponding values of the absolute continuum were determined by interpolation from the data of Allen (1976). The values of the Einstein $A_{u\ell}$ coefficients were taken from Bommier et al. (2006).

From Table~\ref{tab:sol_impli}, one can see that for both $n_{H} \!=\! 10^{15} {\rm cm}^{-3}$ and $10^{16} {\rm cm}^{-3}$, $D^{2}(j_{\ell})$ is comparable to $B_{\ell u} I(\lambda)$. This implies that the $X^{2}\Sigma$ sate of MgH cannot be completely depolarized by collisions. Hence, one has to take into account the lower-level polarization when solving the SEEs to calculate the polarization of observed lines. This is an important result since previously the lower-level polarization was neglected by assuming that it is completely depolarized by collisions (Mohan Rao \& Rangarajan 1999; Asensio Ramos \& Trujillo Bueno 2005; Bommier et al. 2006).

We also consider the Hanle effect due to turbulent magnetic field on the polarization of the MgH ground state, $X^{2}\Sigma$. The Hanle effect is important only if $t_{\rm life}$ of the considered level [$t_{\rm life}\!=\! (B_{\ell u} I(\lambda_{u \ell})^{-1}$ for the ground state] is comparable to $(\omega_L \vert g_{j} \vert)^{-1}$, where $\omega_L \!=\! 8.79 \!\times\! 10^{6}$ B is Larmor angular frequency, with B being the magnetic field strength in gauss.

In Table~\ref{tab:sol_impli}, we display values of $\omega_L \vert g_{j_{\ell}} \vert$ calculated at ${\rm B} \!=\! 10 \; {\rm G}$ and ${\rm B} \!=\! 100  \;  {\rm G}$. One can see that $\omega_L \vert g_{j_{\ell}} \vert \!\gg\! B_{\ell u} I$ in all cases, which implies that for typical values of the photospheric turbulent magnetic field $\sim 10-100$~G the saturation regime of the Hanle
effect on linear polarization of MgH~$X^{2}\Sigma$ is reached.

\section{Conclusion}
We provide (de-)excitation, depolarization, and polarization transfer rates of the MgH $X ^2\Sigma$ state by collisions with neutral hydrogen in its $^2S$ ground state. These rates are important for precise interpretation of MgH blue lines in the SSS. A detailed discussion of the results is presented and general trends of the collisional rates are given so as to gain some understanding
about the completely unknown role of collisions on the polarization of other molecules. We obtain useful variation laws of the depolarization rates with the temperature and the total angular momentum. Important solar implications of our findings are pointed out.

\section*{Acknowledgements}
This research work was funded by the Institutional Fund Projects under grant No. (IFPHI-179-130-2020). Therefore, authors gratefully acknowledge technical and financial support
from the Ministry of Education and King Abdulaziz University, DSR, Jeddah, Saudi Arabia.


\end{document}